%% file: susy09HiggsBounds_arXiv.tex
\documentclass[
    ,final            
  ]
  {aipproc}

\layoutstyle{6x9}

\newcommand{\TOT}{{\mathrm{tot}}}
\newcommand{\HIGGS}{\mathrm{Higgs}}
\newcommand{\BR}{\mathrm{BR}}
\newcommand{\MOD}{{\mathrm{model}}}
\newcommand{\REF}{{\mathrm{ref}}}
\newcommand{\SM}{\mathrm{SM}}
\newcommand{\OBS}{\mathrm{obs}}
\newcommand{\EXPEC}{\mathrm{expec}}

\newcommand{\size}{\footnotesize}

\begin{document}

\input{susy09HiggsBounds_titlepage}

\input{susy09HiggsBounds_main}

\end{document}

%% file: susy09HiggsBounds_titlepage.tex
\thispagestyle{empty}
\setcounter{page}{0}
\def\thefootnote{\fnsymbol{footnote}}

\begin{flushright}
DCPT/09/146\\
FREIBURG-PHENO-09-05\\
IPPP/09/73\\
arXiv:0909.4664 [hep-ph]
\end{flushright}

\vspace{1cm}

\begin{center}

{\fontsize{15}{1} 
\sc {\bf New HiggsBounds from LEP and the Tevatron}}
\footnote{talk given by S.H.\ at the {\em SUSY\,09}, 
June 2009, Boston, USA}

\vspace{1cm}

{\sc 
P.~Bechtle$^1$
\footnote{
email: bechtle@mail.desy.de
}%
, O.~Brein$^2$
\footnote{
email: oliver.brein@physik.uni-freiburg.de
}%
, S.~Heinemeyer$^3$
\footnote{
email: Sven.Heinemeyer@cern.ch
}%
,\\[.5em] G.~Weiglein$^4$
\footnote{
email: Georg.Weiglein@durham.ac.uk
}%
~and K.~Williams$^5$
\footnote{
email: k.e.williams@dunelm.org.uk
}
}

\vspace*{1cm}

$^1$DESY, Notkestrasse 85, 22607 Hamburg, Germany

\vspace{0.1em}

$^2$Physikalisches Institut,
Albert-Ludwigs-Universit\"at Freiburg,
79106 Freiburg, Germany

\vspace{0.1em}

$^3$Instituto de F\'isica de Cantabria (CSIC-UC), Santander,  Spain

\vspace{0.1em}

$^4$Institute for Particle Physics Phenomenology,
Durham University, Durham, DH1 3LE, UK

\vspace{0.1em}

$^5$
Bethe Center for Theoretical Physics, 
Physikalisches Institut der
Universit\"at Bonn,
Nussallee 12, 53115 Bonn, Germany
\end{center}

\vspace*{0.2cm}

\begin{center} {\bf Abstract} \end{center}
\input{susy09HiggsBounds_abstract}

\def\thefootnote{\arabic{footnote}}
\setcounter{footnote}{0}

\newpage

%% file: susy09HiggsBounds_abstract.tex
We review the program {\tt HiggsBounds} that
tests theoretical predictions
of models with arbitrary Higgs sectors against the exclusion bounds
obtained from the Higgs searches at LEP and the Tevatron. 
We explicitly list the bounds that have been added after the first
release of {\tt HiggsBounds}.

%% file: susy09HiggsBounds_main.tex
\title{New HiggsBounds from LEP and the Tevatron}


\author{P.~Bechtle}{
  address={DESY, Notkestrasse 85, 22607 Hamburg, Germany}
}

\author{O.~Brein}{
  address={Physikalisches Institut,
Albert-Ludwigs-Universit\"at Freiburg,
79106 Freiburg, Germany}
}

\author{S.~Heinemeyer}{
  address={Instituto de F\'isica de Cantabria (CSIC-UC), Santander, Spain}
}

\author{G.~Weiglein}{
  address={Institute for Particle Physics Phenomenology,
Durham University, Durham, DH1 3LE, UK}
}

\author{K.~Williams}{
  address={Bethe Center for Theoretical Physics, 
Physikalisches Institut der
Universit\"at Bonn,
Nussallee 12, 53115 Bonn, Germany}
}

\begin{abstract}
\input{susy09HiggsBounds_abstract}
\end{abstract}

\maketitle


\section{Introduction}

The search for Higgs bosons is a major cornerstone
of the physics programs of past, present and future 
high energy colliders. The LEP and Tevatron 
experiments, in particular, turned the non-observation
of Higgs bosons into constraints on the Higgs 
sector, which can be very useful in reducing the available parameter
space of particle physics models. Such constraints will continue 
to be important far into the LHC era as they will need to be taken into
account in the interpretation of any new physics scenario.

The constraints are provided by experiments in the form
of limits on cross sections of individual signal topologies (such as
$e^+e^-\to h_iZ\to b\bar{b}Z$ or $p\bar p\to h_i Z\to b\bar b l^+l^-$) 
or in the form of combined limits for a specific model, such as the
SM. 
The latter type of analyses include detailed knowledge of the overlap
between the individual experimental searches, and therefore have a high
sensitivity, whereas the former can be used to test a wide class of models.

Comparing the predictions of a particular model with the existing
experimental bounds on the various search topologies can be
quite a tedious task as it involves the implementation of experimental
results that are distributed over many different publications and 
combining these results requires a procedure to ensure the correct
statistical interpretation of the exclusion bounds obtained on the
parameter space of the model.

The program {\tt HiggsBounds}~\cite{higgsbounds} is a tool
designed to facilitate the above task so that wide classes of models can
easily be checked against the state-of-the-art results from Higgs
searches. 
This should be useful for applications in Higgs phenomenology
and model building (see, e.g., \cite{diffH,Master3}). 
{\tt HiggsBounds} takes theoretical Higgs sector predictions,
e.g.\ for a particular parameter scenario of a model beyond the SM,
as input. It determines which Higgs search analysis
has the highest exclusion power according to a list of expected
exclusion limits from LEP and the Tevatron (an expected limit 
corresponds to the bound that one would obtain in the hypothetical
case of an observed distribution that agrees precisely with the
background expectation).
In order to ensure the correct statistical
interpretation of the obtained exclusion bound as a 95\% CL, the
comparison of the model with the experimental limits
has to be restricted to the single channel that possesses the 
highest statistical sensitivity. 
For this channel, the program then compares the theoretical prediction 
for the Higgs production cross section 
times decay branching ratio
with the actual experimental limit and determines whether or not
the considered parameter point of the model is excluded at 95\% CL.


\section{The Code {\tt HiggsBounds}}

The code roughly works as follows.
The user provides the Higgs sector predictions of the model (point in
the parameter space) under
consideration. For each neutral Higgs boson 
$h_i \; (i=1,\ldots,n_\HIGGS)$ in the model, 
this will usually include the mass, total decay width, branching ratios
and Higgs production cross sections: 
\begin{equation}
\label{basic input}
M_{h_i} \,, 
\Gamma_{\TOT}(h_i)\,,  
\BR_\MOD(h_i\to ...)\,, 
 \sigma_\MOD(P) / \sigma_{\REF}(P)\, .
\end{equation}
Where it exists, $\sigma^\SM(P)$ is used as the reference cross
section. Variations on this input format are offered. As an example, the
branching ratios and production cross section can be replaced by the
effective couplings of the Higgs boson(s) to SM particles. More 
details are given in the {\tt HiggsBounds} manual~\cite{higgsbounds}. 

A complete list of the experimental analyses included in the first release of 
{\tt HiggsBounds} is given in \cite{higgsbounds}. 
The included results from LEP and the Tevatron consist of
tables of expected (based on MC simulations with no signal) and
observed 95\% CL cross section limits, with a variety of
normalizations. The set mainly consists of analyses for which
model-independent limits were published. 
However, we also included some dedicated analyses carried out for the
case of the SM. These analyses are only taken into account as a possible
exclusion bound 
if the Higgs boson in question would appear sufficiently `SM-like' to
this analysis. Roughly speaking, this requires that the ratios of all
involved couplings to the SM couplings are approximately equal.

For each Higgs process $X$ (here, we treat each combination of Higgs
bosons in each experimental analysis as a separate $X$), 
{\tt HiggsBounds} uses the input to calculate the quantity $Q_\MOD(X)$,
which, up to a normalization factor, is the predicted cross section
times branching ratio for $X$. In order to ensure the correct
statistical interpretation of the results, 
it is crucial to only consider the experimentally observed limit for one
particular $X$. Therefore, {\tt HiggsBounds} must first determine $X_0$,
which is defined as the process $X$ with the highest statistical
sensitivity for the model point under consideration. In order to do
this, the program uses the tables of expected experimental limits to
obtain a quantity $Q_\EXPEC$ corresponding to each $X$. The process with
the largest value of $Q_\MOD/Q_\EXPEC$ is chosen as $X_0$. 
{\tt HiggsBounds} then determines a value for $Q_\OBS$ for this process
$X_0$, using the appropriate table of experimentally observed limits. If 
\begin{equation}
Q_\MOD(X_0) / Q_\OBS(X_0) > 1 \,,
\label{eq:modvsobs}
\end{equation} 
{\tt HiggsBounds} concludes that this particular parameter point is
excluded at 95 \% CL. 

\medskip
The {\tt HiggsBounds} package (current version {\tt 1.2.0}) 
can be obtained from\\[.5em]
\centerline{{\tt http://www.ippp.dur.ac.uk/HiggsBounds}}\\[.5em]
The code has both a Fortran 77 and Fortran 90 version. It can be
operated in a command line mode that can process
input files in a variety of formats, as a subroutine suitable for
inclusion in user applications, and as an online version, available 
at its home page.
The package includes sample programs which demonstrate how {\tt HiggsBounds}
can be used in conjunction with the widely used MSSM Higgs sector programs 
{\tt FeynHiggs}~\cite{feynhiggs,mhiggslong,mhiggsAEC,mhcMSSMlong} and
{\tt CPsuperH}~\cite{cpsh}.


\section{Newly implemented and updated bounds}

After the first release of {\tt HiggsBounds} more search channels from
LEP~\cite{LEPHiggsSM,LEPHiggsMSSM,LEPflavindep,LEPHgaga} and the 
Tevatron~\cite{CDFHVbbMET,D0HVbbMET,CDFHWW,D0hbbtautau,D0hgaga} have
been implemented on top of what is described in~\cite{higgsbounds}.
These newly implemented bounds are summarized in Tab.~\ref{tab:new}.

\renewcommand{\arraystretch}{1.5}
\begin{table}[htb!]
\begin{tabular}{ll}
\hline
{\size $e^+e^-\to h_k \, Z\to X + Z$ }~\cite{LEPflavindep} &
{\size $e^+e^-\to h_k \, Z \to \gamma\gamma\; Z$}~\cite{LEPHgaga} 
\\
{\size $p\bar p \to h_k V \to b\bar b$ $+$ missing $E_T$
                                              \cite{CDFHVbbMET,D0HVbbMET} } &
{\size $p \bar p \to H + X \to WW + X$~\cite{CDFHWW,D0HWW}}\\
{\size $p \bar p \to h_k b \to \tau^+\tau^- b$~\cite{D0hbbtautau}} &
{\size $p \bar p \to H + X$ 
 (SM combined)}~\cite{CDFD0HWW,CDFhxx,D0hxxSM,CDFhxxSM} \\
\hline
\end{tabular}
\caption{Summary of newly implemented bounds into {\tt HiggsBounds}
after its first release~\cite{higgsbounds}. $h_k$ denotes a generic
neutral Higgs boson, while $H$ denotes a SM-like Higgs
boson. 
}
\label{tab:new}
\end{table}
\renewcommand{\arraystretch}{1.0}

The code is constantly kept up-to-date by the inclusion of updated results
published by the Tevatron experiments. This is illustrated in
Tab.~\ref{tab:update}, where we list the search channels that have been
updated since the initial release.

\renewcommand{\arraystretch}{1.5}
\begin{table}[htb!]
\begin{tabular}{ll}
\hline
{\size $p \bar p \to h_k\,Z \to b \bar b \, ll$}~\cite{D0hZbb,CDFhZbb,CDFHSMcomb} &
{\size $p \bar p \to h_k\,W \to b \bar b \, l \nu_l$}~\cite{D0hWbb,CDFhWbb,D0HW,CDFHW}
\\
{\size $p \bar p \to h_k\,W \to WWW \to l \nu_l \, l \nu_l + X$}~\cite{D0hWWW,CDFhWWW} &
{\size $p \bar p \to H + X \to \gamma\gamma + X$}~\cite{D0hgaga,D0Hgaga} 
\\
{\size $p \bar p \to h_k \to \tau^+\tau^-$}~\cite{CDFD0htautau,CDFhtautau,D0htautau} &
\\
\hline
\end{tabular}
\caption{Summary of Tevatron bounds that have been updated within 
{\tt HiggsBounds} 
after its first release~\cite{higgsbounds}. $h_k$ denotes a generic
neutral Higgs boson, while $H$ denotes a SM-like Higgs
boson. 
}
\label{tab:update}
\end{table}
\renewcommand{\arraystretch}{1.0}

Still missing are the bounds on charged Higgs bosons from both LEP and
the Tevatron. They will be 
implemented into {\tt HiggsBounds} in the near future.


\begin{theacknowledgments}
We are grateful for the valuable assistance of A. Read, P. Igo-Kemenes,
M.~Owen, T.~Junk, M.~Herndon and S. Pagan Griso. 
This work has been supported 
in part by the European Community's Marie-Curie Research
Training Network under contract MRTN-CT-2006-035505
`Tools and Precision Calculations for Physics Discoveries at Colliders'
(HEPTOOLS). P.B.\ was partially supported by 
the Helmholtz Young Investogator Grant VH-NG-303 and the DFG
Collaborative Research Center SFB~676.
\end{theacknowledgments}

\bibliographystyle{aipprocl} 